\documentclass[aps,prl,reprint,superscriptaddress]{revtex4-1}

\usepackage{amsmath}    
\usepackage{graphicx}   
\usepackage{verbatim}   
\usepackage{color}      
\usepackage{subfigure}  
\usepackage{hyperref}   
\usepackage[version=4]{mhchem} 
\usepackage{blindtext} 
\usepackage{siunitx}

\usepackage{enumitem}

\begin{document}

\title{Anomalous phonon life time shortening in paramagnetic CrN caused by magneto-lattice coupling:  A combined spin and \textit{ab initio} molecular dynamics study}
\author{Irina Stockem}
\email{irina.stockem@liu.se}
\affiliation{Department of Physics, Chemistry, and Biology (IFM), Link\"oping University, 581 83  Link\"oping, Sweden}
\affiliation{Max-Planck-Institut f\"ur Eisenforschung GmbH, 402 37 D\"usseldorf, Germany}
\author{Anders Bergman}
\affiliation{Maison de la Simulation, USR 3441, CEA-CNRS-INRIA-Universit\'e Paris-Sud-Universit\'e de Versailles, 91191 Gif-sur-Yvette, France }
\affiliation{L\_Sim, INAC-MEM, CEA, 38000 Grenoble, France}
\author{Albert Glensk}
\affiliation{Max-Planck-Institut f\"ur Eisenforschung GmbH, 402 37 D\"usseldorf, Germany}
\author{Tilmann Hickel}
\affiliation{Max-Planck-Institut f\"ur Eisenforschung GmbH, 402 37 D\"usseldorf, Germany}
\author{Fritz K\"ormann}
\affiliation{Max-Planck-Institut f\"ur Eisenforschung GmbH, 402 37 D\"usseldorf, Germany}
\affiliation{Department of Materials Science and Engineering, Delft University of Technology, Mekelweg 2, 2628 CD Delft, The Netherlands}
\author{Blazej Grabowski}
\affiliation{Max-Planck-Institut f\"ur Eisenforschung GmbH, 402 37 D\"usseldorf, Germany}
\author{J\"org Neugebauer}
\affiliation{Max-Planck-Institut f\"ur Eisenforschung GmbH, 402 37 D\"usseldorf, Germany}
\author{Bj\"orn Alling}
\affiliation{Department of Physics, Chemistry, and Biology (IFM), Link\"oping University, 581 83  Link\"oping, Sweden}
\affiliation{Max-Planck-Institut f\"ur Eisenforschung GmbH, 402 37 D\"usseldorf, Germany}

\date{\today}

\begin{abstract}

We study the mutual coupling of spin fluctuations and lattice vibrations in paramagnetic CrN by combining atomistic spin dynamics and \textit{ab initio}  molecular dynamics. The two degrees of freedom are dynamically coupled leading to non-adiabatic effects. Those effects suppress the phonon life times at low temperature compared to an adiabatic approach. The here identified dynamic coupling provides an explanation for the experimentally observed unexpected temperature dependence of the thermal conductivity of magnetic semiconductors above the magnetic ordering temperature.

\end{abstract}

\maketitle
 
\begin{figure}[h!btp]
 \centering
   \includegraphics[width=1.0\linewidth]{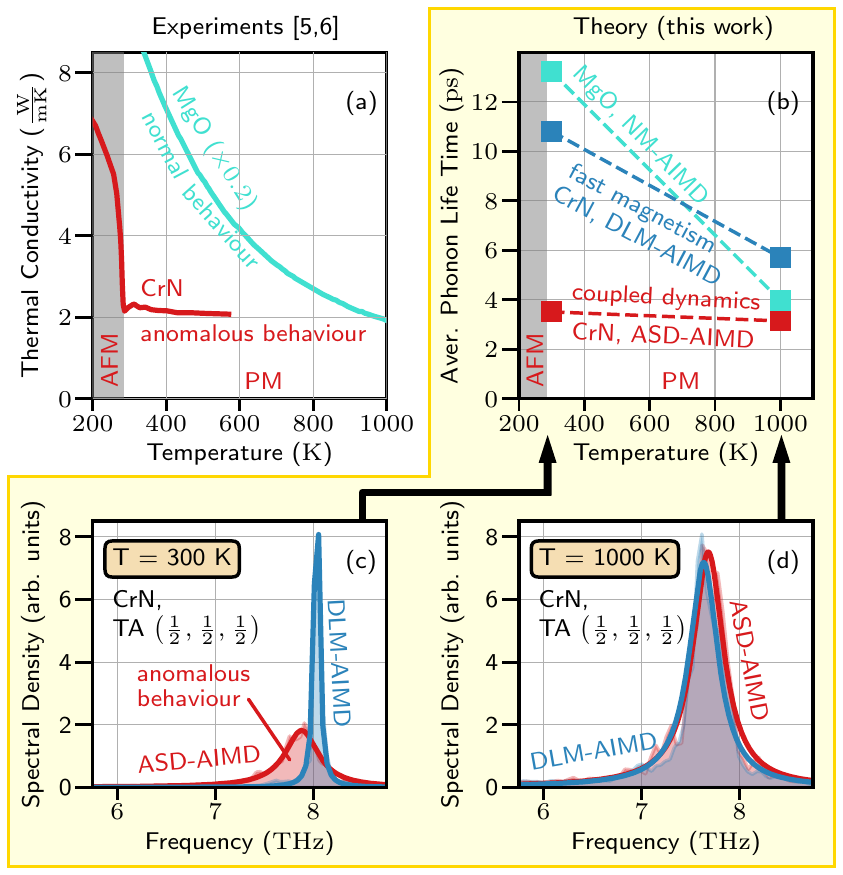} 
   \caption{(Color online) (a) Thermal conductivities of CrN~\cite{Jankovsky2014} and MgO~\cite{Powell1968} (b) compared to averaged phonon lifetimes obtained by DLM-AIMD, ASD-AIMD and nonmagnetic AIMD. (c) CrN spectral density of the transversal acoustic phonon at the L-point at $T=\SI{300}{K}$ (d) and $T= \SI{1000}{K}$. Shaded areas indicate the corresponding non-fitted spectral densities.}
   \label{fig:Phonon}
\end{figure}

At elevated temperatures magnetic materials are subject to both spin and lattice excitations. Magnon and phonon spectra, as well as magnetic order-disorder and melting transitions are often analysed separately even though in principle they are distinctly coupled~\cite{Fransson2017}. As an example, the phonon dominated thermal conductivity of magnetic semiconductors \ce{YMnO3}, \ce{LuMnO3} and \ce{ScMnO3}~\cite{Sharma2004}, as well as CrN~\cite{Quintela2009, Tomes2011, Jankovsky2014} shows dramatically different temperature dependences below and above the magnetic transition temperature. Particularly above the magnetic ordering temperature, the thermal conductivity differs from the temperature dependence of nonmagnetic semiconductors like \ce{MgO} \cite{Powell1968}. While the phonon based thermal conductivity of nonmagnetic materials typically declines by $1/T$ at high temperatures due to increasing phonon-phonon interactions (turquois line in Fig. \ref{fig:Phonon} (a)), the thermal conductivity of magnetic semiconductors such as CrN remains nearly constant above the magnetic transition (red line in Fig. \ref{fig:Phonon} (a)). So far a well-supported explanation for this behavior has been lacking. We show here that in the high temperature paramagnetic phase, non-adiabatic coupling between phonons and spins strongly affects the phonon life times (Fig. \ref{fig:Phonon} (b)). The phonon life time is directly correlated to the thermal conductivity. Our findings give an atomistic explanation and support the suggestion in the experimental work of Sharma et al.~\cite{Sharma2004}, who proposed a dynamical spin-phonon scattering mechanism of acoustic phonons by short-ranged spin fluctuations.

\begin{figure*}[h!tbp]
   \centering
   \includegraphics[width=1.0\textwidth] {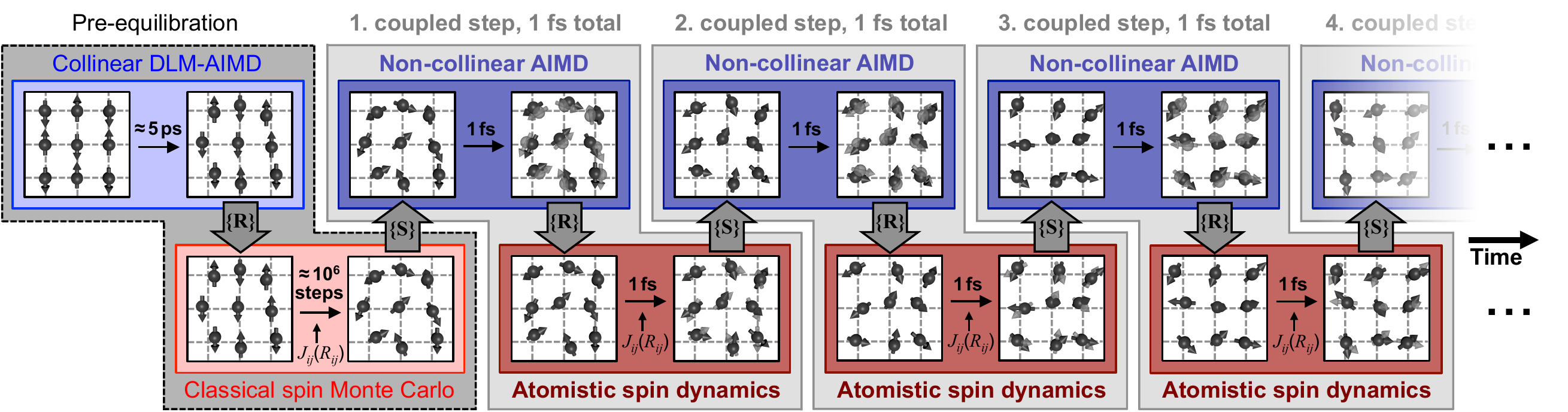} 
   \caption{(Color online) Representation of the combined ASD-AIMD approach: After a pre-equilibration, a subsequent scheme of ASD and AIMD runs are performed. At each step, the actual spin state $\{\mathbf{S}\}$ and the actual atomic positions $\{\mathbf{R}\}$ serve as input of the following AIMD and ASD calculation, respectivly. Within each coupled step the atoms and spins propagate for $\SI{1}{fs}$ in total.}
   \label{fig:ASDAIMD}
\end{figure*}

Above the critical Curie or N\'eel temperature the paramagnetic state is characterized by a lack of both global magnetization and long range order between the magnetic moments. Nevertheless, local moments with a degree of short-range correlations both in time and space may still exist and influence physical properties~\cite{Rivadulla2009, Alling2010, Wang2012}. Furthermore, the lattice degrees of freedom experience thermal excitations, possibly well beyond the harmonic regime~\cite{Dutta2017, Pradip2016, Glensk2015}. Recently it has been demonstrated by first-principles calculations that the lattice and magnetic excitations are interconnected in the paramagnetic phases of Fe: The interatomic forces that dictate the vibrations and the phonon spectra are distinctly different in the disordered magnetic state as compared to the ferromagnetic ground state~\cite{Koermann2012, Koermann2014}. The local magnetic moments and electronic structure are in turn clearly influenced by the presence of atomic disorder caused by the lattice vibrations~\cite{Alling2016}.

Previous approaches~\cite{Koermann2012, Koermann2014, Steneteg2012} lack, however, an incorporation of the full {\it mutual} coupling between both degrees of freedom. For instance, in the previously proposed spin-space averaging approach~\cite{Koermann2012, Koermann2014}, the atomic fluctuations do not directly impact the magnetic system. In the disordered local moment ab initio molecular dynamics (DLM-AIMD)~\cite{Steneteg2012} approach magnetic short and long range ordering effects are ignored. Properties such as the experimentally observed anomalous thermal conductivity in CrN cannot be interpreted by these methods. 

An alternative would be spin-lattice dynamics (SLD)~\cite{Ma2008, Ma2016, Ma2017, Perera2016, Perera2017}, but efficient implementations require adequate interatomic potentials for the magnetic and atomic degree of freedom.  Constructing these is particularly challenging for magnetic materials at high temperatures. A key prerequisite for such a method development is thus an unbiased and fully first-principles based approach.


In this work we suggest an approach to close the previous simulation gap for magnetic materials by a first-principles based methodology. Our approach combines atomistic spin dynamics~\cite{Antropov1996, Skubic2008} with \textit{ab initio} molecular dynamics (ASD-AIMD) and allows the explicit study of coupled dynamical excitations. Based on this approach we derive magnetic and phonon lifetimes in B1 CrN above the magnetic ordering temperature of $\SI{280}{K}$.

Our ASD-AIMD method is sketched in Fig.~\ref{fig:ASDAIMD}. It is based on an alternating scheme of ASD and AIMD steps:
\begin{itemize}[leftmargin=*]
\item Pre-equilibration: To start with suitable initial atomic positions, adiabatic DLM-AIMD simulations are performed at the temperature of choice. The resulting positions $\{\mathbf{R}\}$ are used to obtain the distance dependent exchange constants $J_{ij}$ between magnetic atoms, which are parametrized prior to the simulation. A reasonable initial spin state $\{\mathbf{S}\}$ is then determined by a Monte Carlo (MC) simulation for these atomic positions and corresponding exchange interactions. After this initiation the main loop of the ASD-AIMD is started.
\item The spin orientations are used for the subsequent AIMD step. There, forces acting on each atom are determined via a spin polarized, non-collinear density functional theory (DFT) calculation with constrained spin directions consistent with the actual orientation of the spins. The obtained forces are used to update the atomic positions by a single-step propagation of $\SI{1}{fs}$.
\item With the new atomic positions, new distance dependent exchange interactions are obtained and used to propagate the magnetic state for $\SI{1}{fs}$ in the ASD simulation of the same coupled ASD-AIMD step. 
\item The alternating AIMD and ASD steps are repeated until a sufficient simulation time has been obtained to ensure convergence of the investigated physical quantities.
\end{itemize}
The propagation of the atoms has been realized by a Born-Oppenheimer type AIMD within VASP~\cite{Kresse1993, Kresse1996, Kresse1996PhysRevB, Kresse1996, Kresse1999, Bloechel1994, Hobbs2000}. The forces acting on the atoms are determined in formally $\SI{0}{K}$ DFT calculations with a constrained magnetic moment approach by Ma and Dudarev~\cite{Ma2015} via Lagrange multipliers of magnitude $\lambda=25$. The actual change of the spin state is calculated by the Landau-Lifshitz-Gilbert equation (LLG) as implemented in UppASD~\cite{Skubic2008, Eriksson2017}
\begin{align} \label{eq:LLG}
\frac{\partial \hat{\mathbf{S}}_i}{\partial t} = &-\frac{\gamma}{1+\alpha^2} \hat{\mathbf{S}}_i \times \left[ \mathbf{H}_\mathrm{eff} + \mathbf{f}_i  \right] \\
&-\gamma \frac{\alpha}{1+\alpha^2}  \hat{\mathbf{S}}_i  \times  \{ \hat{\mathbf{S}}_i  \times \left[ \mathbf{H}_\mathrm{eff} + \mathbf{f}_i \right]  \} \nonumber.
\end{align}
The quantities $\gamma$ and $\alpha$ denote the electron gyromagnetic ratio and the phenomenological damping factor, respectively. The temperature fluctuations are described by an additional stochastic magnetic field $\mathbf{f}_i$. Both parts of Eq.~\ref{eq:LLG} are based on the effective field $\mathbf{H}_\mathrm{eff}= -\frac{1}{m_i}\frac{\partial \mathcal{H}}{\partial \hat{\mathbf{S}}_i}$ with $m_i$ denoting the magnitude of the magnetic moment corresponding to the $i$-th atom. The effective field is determined by the classical Heisenberg exchange interaction between the effective spin of  Cr atoms:
\begin{equation} \label{eq:HeisenbergHamiltonian}
\mathcal{H}=-\sum_{i\neq j} J_{ij}\left({R}_{ij}\right) \hat{\mathbf{S}}_i \hat{\mathbf{S}}_j.
\end{equation}
The lattice vibrations enter the spin dynamics through their effect on the exchange interactions $J_{ij}\left(R_{ij}\right)$. The CrN exchange interactions have been determined in~\cite{Lindmaa2013} and depend mostly on the Cr-Cr atomic pair distances. The range of interactions in paramagnetic CrN was found to be very short and could be well described with only first and second Cr-Cr neighbor terms~\cite{Lindmaa2013, LindmaaMasterThesis}. The parametrization of these interactions are determined by magnetic direct cluster averaging using the same electronic structure framework and approximations as used for our AIMD runs~\cite{Lindmaa2013} (see supplement~\cite{SupplementaryMaterials}).
Within our notation, the magnitudes of the effective magnetic moments $m_i$ and $m_j$ are absorbed into the exchange interactions $J_{ij}\left(R_{ij}\right)$.  

While N carries a negligible magnetic moment, the magnitude of Cr moments in CrN has been observed to be robust around 2.8~$\mu_B$ with respect to both transversal magnetic disorder~\cite{AllingTiCrN2010} and lattice vibrations~\cite{Shulumba2014}. We therefore constrain the transverse, orientational, degree of freedom in the LLG equation only. The absolute magnitude of each moment is allowed to adjust freely within each electronic structure step. As discussed in the supplement, the phenomenological damping factor has a small influence on the spin dynamics in equilibrium simulations of the high temperature phase and the typical value of 0.05 has been chosen. 
We applied our method to CrN to investigate the properties within the paramagnetic phase at $T=\SI{300}{K}$ and $T=\SI{1000}{K}$. Detailed information of DFT related parameters (e.g. k-points, energy cutoff, cell sizes) are given in the supplement.

To analyse the importance of a spin-phonon coupling, we performed DLM-AIMD~\cite{Steneteg2012} calculations employing the same $\SI{1}{fs}$ AIMD time step but with fully random, disordered non-collinear magnetic states changing likewise each $\SI{1}{fs}$. In this way we approximate the adiabatic limit where individual spin states live too short to be "seen" by the atomic vibrations. This can be regarded as a generalization of previous DLM-AIMD works on CrN~\cite{Steneteg2012, Mozafari2016}. It considers non-collinear disorder in the limit of an infinitely fast change of magnetic states. All other details of the DLM-AIMD simulations are identical to the ASD-AIMD runs. 

The most compelling finding is observed for the phonon lifetimes of CrN in the paramagnetic range (see supplement for a description on obtaining phonon lifetimes from MD). In Fig. \ref{fig:Phonon} (c) and (d) we show the power spectral density of the transversal acoustic phonon at the $L$-point as a representative example. There, we compare the coupled ASD-AIMD simulation to the adiabatic-like DLM-AIMD, both at (a) $\SI{300}{K}$, and (b) $\SI{1000}{K}$. We observe small effects on the phonon frequencies (corresponding to the peak positions). However, the full widths at half maximum ($\mathrm{FWHM} = 2 \Gamma_{\rm{ph}}$), which are inversely proportional to the phonon lifetimes ($\tau=\frac{1}{2 \Gamma_{\rm{ph}}}$), are distinctly different with significantly broader distributions for the coupled ASD-AIMD simulation. This effect is particularly apparent at $\SI{300}{K}$ and largely vanishes at $\SI{1000}{K}$. This impressively demonstrates that at $\SI{300}{K}$, above the magnetic transition temperature, the vibrational state is heavily disturbed by the dynamical coupling to the magnetic state without long-range order. In panel (b) of Fig.~\ref{fig:Phonon} we show the average of all obtained phonon lifetimes for CrN simulated with the two methods, as well as for the isostructural nonmagnetic \ce{MgO}. The absolute value at $\SI{300}{K}$ and the temperature trend of the averaged phonon lifetimes are distinctly different between the adiabatic-like DLM-AIMD and the dynamically coupled ASD-AIMD simulations. The adiabatic-like simulation for CrN shows a striking resemblance with the results for nonmagnetic simulations for $\ce{MgO}$. The similar temperature trend indicates that disordered magnetism treated as adiabatically fast on the timescale of the phonons, has little influence on the phonon lifetime and also on their temperature dependence. In contrast, the results for the ASD-AIMD simulation show that when the spin dynamics is treated on an equal footing and coupled with the lattice dynamics it can indeed drastically decrease phonon lifetimes. Our results therefore show that the coupling strength is large at $\SI{300}{K}$ and that it decreases with temperature. Decomposing the phonons into acoustic and optical branches (cf. supplement) we find that the difference between the DLM and ASD treatments originates from effects on the acoustic branches, while the lifetimes of the optical branches are very similar. This may be expected as the optical modes are governed by the interatomic forces of neighboring Cr-N pairs where the N atoms are almost absent of spin polarization and in particular fully absent of spin dynamics.

Since the phonon lifetimes are are directly related to the lattice thermal conductivity, these findings provide an intuitive explanation of the constant \cite{Tomes2011, Jankovsky2014} or even increasing~\cite{Quintela2009} thermal conductivity with temperature in the paramagnetic state of CrN, as shown by the measured data from \cite{Jankovsky2014} in Fig. \ref{fig:Phonon} (a). Normally phonon life times reduce with increasing temperature due to increased phonon-phonon scattering. As discussed above, the dynamic coupling between spin dynamics and lattice vibrations has the opposite temperature dependence - it becomes weaker with increasing temperature. The reduction of the acoustic phonon life times is strong close to the N\'eel temperature compared to the case of phonon-phonon interaction only. In contrast, at high temperatures the coupled dynamics has little impact on the phonon life times. As a consequence of this compensating behavior the phonon life time becomes almost temperature independent.

\begin{figure}[t!bp]
 \centering
   \includegraphics[width=1\linewidth]{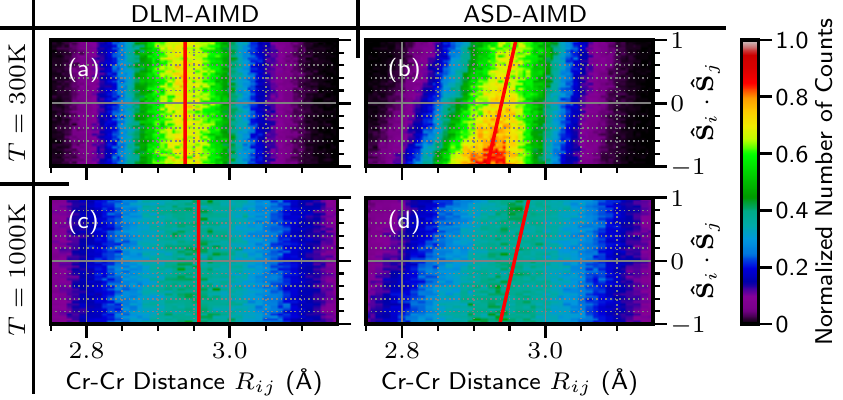} 
   \caption{(Color online) Two dimensional probability diagrams of Cr-Cr nearest neighbor pairs with respect to their spin orientation. The spin orientation is given by the dot products of the two spin vectors in a pair (1 (-1) corresponds to the ferro- (antiferro-) magnetic coupling). The diagrams are plotted for different temperatures and methods: (a) DLM-AIMD at $T=\SI{300}{K}$, (b) ASD-AIMD at $T= \SI{300}{K}$, (c) DLM-AIMD at $T= \SI{1000}{K}$ and (d) ASD-AIMD at $T= \SI{1000}{K}$. The red lines show the linear regression between distances and spin orientations.}
   \label{fig:DotDist}
\end{figure}

\begin{figure}[t!bp]
   \centering
         \includegraphics[width=1.0\linewidth]{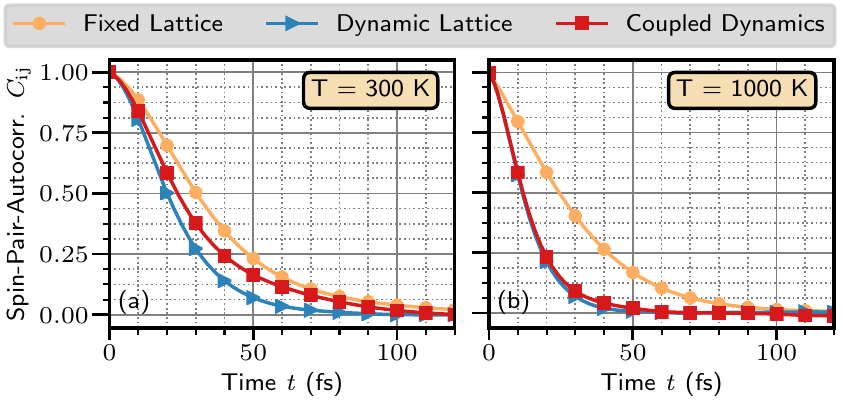}
   \caption{(Color online) Spin pair autocorrelation functions of CrN at $T= \SI{300}{K}$ (a) and $T= \SI{1000}{K}$ (b) obtained from ASD (fixed lattice), ASD on a pre-calculated DLM-AIMD vibrating lattice (dynamic lattice) and ASD-AIMD (coupled dynamics). }
   \label{fig:Correlation}
\end{figure}

To obtain a detailed understanding of  the nature of the coupling effect and its change with temperature, we study i) the correlation between the momentaneous interatomic distance and spin state (Fig. \ref{fig:DotDist}) and ii) the influence of the lattice vibrations on the magnetic short range order autocorrelations (Fig. \ref{fig:Correlation}). The spin dot product $\mathbf{\hat{S}_i} \cdot \mathbf{\hat{S}_j}$ as a function of  Cr-Cr nearest neighbor pair distance $R_{ij}$ is shown in Fig. \ref{fig:DotDist} as a histogram for every AIMD snapshot. The histogram represents the probability to find a pair with a specific distance and spin orientation. Including dynamic coupling (Fig. \ref{fig:DotDist} (b) and (d)) a distinct correlation between distances and spin orientations is observed. Specifically the AFM oriented pairs have a tendency for shorter distances while FM oriented pairs have a tendency for longer distances. The higher probability at antiferromagnetic orientations (indicated by the red color) implies that the system shows  substantial antiferromagnetic short range order (SRO) at $\SI{300}{K}$. At a higher temperature of $\SI{1000}{K}$ (Fig. \ref{fig:DotDist} (d)) local spin order largely disappears and the correlation between pair separation and spin orientation is less pronounced relative to the larger total spread in pair distances. As expected, when not taking the dynamic coupling into account (Fig. \ref{fig:DotDist} (a) and (c)) no correlation is observed. In order to verify that the spin-lattice correlation is indeed due to the dynamical coupling and not due to spatial magnetic SRO we replace the spin-dynamic step in our approach by Monte Carlo (MC) calculations independent from each other but including magnetic SRO. The comparison clearly shows, Fig. 5 and 6 in the supplement, a magnetic SRO at $300K$ without impact on interatomic distances or suppression of phonon life times, proving the dynamical origin of the latter effects. 

Having studied the impact of the spin dynamics on the dynamics of the nuclei we now study the opposite effect, i.e., the impact of the atomic vibrations on spin dynamics. For this reason we consider three cases. In the first case we keep the atoms fixed on their equilibrium positions, i.e., by construction there is no atomic dynamics. Second, we switch on atomic dynamics but we do not couple it to the spin dynamics. Nevertheless, the spin dynamics is influenced by the distances between magnetic atoms. Third, we consider the fully coupled case. As shown in Fig. 4 including atomic dynamics always reduces the spin correlation time. The magnitude of this decrease depends significantly on the temperature: While at low temperatures shortening of spin correlation times is modest it becomes large at high temperatures. What is interesting is the qualitatively different impact of the dynamical coupling. While at high temperatures there is no visible difference between coupled and non-coupled atom dynamics, at low temperatures there are clear differences. Including dynamical coupling between spin and lattice increases the spin correlation times, i.e., spin scattering at the vibrating atoms becomes less efficient.

The weaker magneto-lattice coupling at $\SI{1000}{K}$ than at $\SI{300}{K}$ can be understood by two effects: First, the even shorter life times of the magnetic state at $\SI{1000}{K}$, seen in Fig. \ref{fig:Correlation}, brings the system towards an adiabatic de-coupling. Second, the higher amplitude of atomic displacements, seen in the distribution width in Fig. \ref{fig:DotDist}, corresponds to an increased vibrational energy scale not matched by an increase in the energy scale of the magnetic fluctuations that already at $\SI{300}{K}$ is in the disordered paramagnetic regime.


In conclusion we have proposed a method that allows to simulate the high temperature paramagnetic phase of magnetic materials by combining \textit{ab initio} parametrized atomistic spin dynamics with \textit{ab initio} molecular dynamics. The approach enables studies of the dynamical coupling of spin and lattice excitations and its impact on materials properties. For the model system studied, CrN, we find that the dynamics of the spin system severely impacts atomic lattice vibrations and quantities related to it such as phonon life times. The dependence is mutual and the magnetic state and its dynamic are influenced by the lattice vibrations. At temperatures slightly above the transition temperature, the dynamical coupling, which is accessible by our approach, is found to reduce significantly the phonon lifetimes of the acoustic modes. In contrast, at high temperatures ($\SI{1000}{K}$) well above the N\'eel temperature, the dynamical coupling and its impact on phonon life times are largely reduced. Thus, the strong magneto-lattice scattering identified in this study and the fact that it is found to decrease with increasing temperature are consistent with experimental observations. Since these mechanisms are generic we expect them to apply for a wide range of magnetic semiconductors.  

\begin{acknowledgments}
A. B. acknowledges the CEA-Enhanced Eurotalents program, co-funded by FP7 Marie Sk{\l}odowska-Curie COFUND Programme (Grant Agreement no 600382), T. H. acknowledges the priority programme SPP1599 "Ferroic cooling" (Grant No. HI1300/6-2), F.K. acknowledges funding by the Netherlands Organisation for Scientific Research (NWO) under the VIDI research programme with project number 15707.  B.G. acknowledges funding by the European Research Council (ERC) under the EU's Horizon 2020 Research and Innovation Programme (Grant No. 639211). B.A. acknowledges financial support by the Swedish Research Council (VR) through the International Career Grant No. 330-2014-6336 and by Marie Sk{\l}odowska Curie Actions, Cofund, Project INCA 600398, the Swedish Government Strategic Research Area in Materials Science on Functional Materials at Link\"{o}ping University (Faculty Grant SFOMatLiU No 2009 00971), as well as support from the Swedish Foundation for Strategic Research through the Future Research Leaders 6 program. Per Eklund is acknowledged for useful discussions.
\end{acknowledgments}


%

\bibliographystyle{apsrev4-1}
\end{document}